\documentclass[aps,twocolumn,groupedaddress]{revtex4}

\usepackage{graphicx}

\newcommand{ \be }{\begin{equation}}
\newcommand{ \ee }{\end{equation}}
\newcommand{ \bea }{\begin{eqnarray}}
\newcommand{ \eea }{\end{eqnarray}}

\begin{document}
\voffset=0.5 in

\title{Two-particle correlations in Au+Au collisions 
at $\sqrt{s_{NN}} = 130$ GeV}


\author{Fabrice Reti\`ere\email[]{fgretiere@lbl.gov} for the STAR collaboration}
%
\affiliation{Lawrence Berkeley National Laboratory, Berkeley, CA 94720, USA}


\begin{abstract}

  We present preliminary results on two new two-particle correlation analyses 
of Au+Au collisions 
at $\sqrt{s_{NN}} = 130 GeV$ performed by the STAR collaboration. 
Two-pion interferometry with respect to the reaction plane
 shows that the pion  source can be described as an ellipse extended out of the 
reaction plane. Analysing the pion-kaon correlation
functions, we show that, on average, pions and kaons are not emitted at
the same position 
and/or time. Both measurements are interpreted in the so-called blast
wave model framework whose main feature is a strong flow.

\end{abstract}


\maketitle


\section{Introduction}

  Two-particle interferometry is a sensitive probe of the 
space-time  geometry of the particle emitting source.  Using this technique,
pion source sizes have been measured at every available relativistic heavy ion 
beam 
energy which led to the excitation function reported in \cite{STARHbt}.
Two-particle correlation also offers a large variety of observables
combining different particle species. For example, two-kaon and two-proton 
correlation 
functions have been measured at the CERN SPS 
~\cite{NA44Kaon,NA49Proton}.  In this paper, we present preliminary results 
from two-particle correlation analyses of
Au+Au collisions  at $\sqrt{s_{NN}} = 130$ GeV produced by the Relativistic
Heavy Ion Collider (RHIC) at Brookhaven National Laboratory.

Two-particle correlations are studied by
constructing correlation functions  : 
$C_{2}(k^{*}) = A(k^{*}) /  B(k^{*})$ 
with $k^{*} = \frac{p_{1}^{*} - p_{2}^{*}}{2}$ the momentum of one particle in the 
rest frame of pair,
$A(k^{*})$ the distribution of $k^{*}$ for pairs of particles from the same event, and 
$B(k^{*})$ the $k^{*}$
distribution for pairs of particle from different events. Note that for
pairs of particles having the same mass, 
$Q_{inv} =  \sqrt{(p_{1}-p_{2})^{2} -(E_{1}-E_{2})^{2}}= 2 k^{*}$, where $Q_{inv}$ is 
the invariant
relative momentum. For pairs of particles having different masses, this relationship 
doesn't hold and $k^{*}$ is the only relevant variable.

Transverse mass spectra~\cite{NuQM} 
and elliptic flow analysis~\cite{STARFlow} suggest that there is a strong 
collective motion of the
particles in the Au+Au collisions  at $\sqrt{s_{NN}} = 130$ GeV.
Such flow introduces a strong correlation between the position
and the momentum of the emitted particles, which reduces the apparent source 
size. 
It could explain why the analysis of two-pion correlation functions published by 
STAR~\cite{STARHbt} shows no anomalously large pion source size which would 
have indicated the formation of the quark gluon plasma.  
Transverse flow must be accounted for in order to interpret the source 
size extracted from two-particle correlation analysis. We will introduce
the so called "extended blast wave model" in order to investigate the effect of 
flow.

  We present two new analyses that are potentially more sensitive to
flow effects. They both provide new independent insights 
on how the system looks at kinetic freeze-out, i.e. when the particles 
decouple.   

\section{The extended blast wave model}

  The extended blast wave model allows one to combine  transverse mass spectra,  
elliptic flow 
and two-particle correlation analysis within a single framework. It is a simple 
parametrization of
the system at kinetic freeze-out in terms of temperature, transverse flow, and
source transverse radius. This approach focuses on the transverse
plane, ignoring the longitudinal direction, i.e., the direction
parallel to the beam.
The blast wave model was introduced in ~\cite{STARFlow} to interpret
the identified particle elliptic flow from STAR. In this model, particles are emitted
from a infinitely thin shell. To reproduce the pion source size measured,
the model had to be extended to a filled cylinder. Such a parametrisation
had been used for several years to interpret transverse mass spectra 
~\cite{Sinyukov,SSH} assuming a anzymuthally symetric system.

  In this extended blast wave framework, the probability of emitting a particle at
a given space-time X and energy-momentum P is :
\bea
f(X,P) & = & f(r, \phi_s, t, p_T, \phi_p, m)  \nonumber \\
            & = & \Theta(1 - \frac{(r \cdot cos(\phi_s) )^2}{R_x^2} -
                             \frac{(r \cdot sin(\phi_s) )^2}{R_y^2})  \cdot \nonumber \\
            &  & K_1(\beta(r,\phi_s,m_T)) \cdot  e^{\alpha(r,\phi_s,p_T)  
                       \cdot \cos(\phi_b-\phi_p)} \cdot  \nonumber \\
            &  & e^{-t^2/\tau^2}
\eea
This equation can be understood separating it in 3 parts:  \\
- The step function, $\Theta$ confines the system inside a filled ellipse. $\phi_s$ 
($\phi_p$) is 
the  spatial (momentum) azymuthal angle ; $\phi_s=0$ for particle emitted in the 
reaction 
plane. $R_x$ and $R_y$ are the maximum radii of the system in plane and out of 
plane, respsectively.  To interpret the identified particle elliptic flow
\cite{STARFlow}, the variable $s_2$ was introduced as a parameter that
quantifies the spatial anisotropy of the particle emission in non-central 
collisions. We find, when describing the system by an ellipse, that  
$s_2 \approx \frac{1}{2} \frac{\eta^3-1}{\eta^3+1}$ with $\eta = R_y / R_x$. 
When $s_2 > 0$, $R_y > R_x$ which means that the source is described by
an ellipse extended out of the reaction plane. \\
- The second term of the equation express the hydro-like behavior of the 
system.  The expression was derived in \cite{SSH} where the following functions 
are used :
$\alpha(r,\phi_s,p_T) = \frac{p_T}{T} \sinh(\rho(r,\phi_s))$ and
$\beta(r,\phi_s,m_T) = \frac{m_T}{T} \cosh(\rho(r,\phi_s))$.
$\rho$ is the transverse rapidity boost that the particles feel at a given emission 
point. In order to  describe non-azymuthally symetric systems, we have modified 
its original expression, to include its dependence on the azymuthal angle :  
$\rho(r,\phi_s) = \frac{3}{2}\frac{\tilde{r}}{R_y}(\rho_0 + \rho_a \cos(2\phi_s))$, 
where $\tilde{r}=\sqrt{(r \cdot sin(\phi_s))^2+\eta^2 (r \cdot cos(\phi_s))^2}$ is the 
"elliptical radius".  I.e. all points on an ellipse have a constant $\tilde{r}$. 
$\rho_a > 0$  means that  there is a stronger boost in the reaction 
plane than out of it.  Since in an ellipse the radial vector and the vector
normal to the surface of the ellipse are not aligned we introduce the
variable $\phi_b = tan^{-1}(\tan(\phi_s)/\eta^2)$. Indeed, in the Cooper Frye
prescription~\cite{CooperFrye} that the blast wave model follows, the transverse 
boost that the particles acquire has to be normal to the freeze-out surface. \\
- The third term of the equation accounts for the duration of particle emission
using a single parameter $\tau$.
This term is particulary
important when interpreting the pion source size. It has no effect, however, on
the transverse mass spectra and elliptic flow. 

In the azymuthally symetric case, this formula simplifies to :
\bea
f(X,P) & = & f(r, \phi_s, t, p_T, \phi_p, m)  \nonumber \\
            & = & \Theta(R-r) \cdot \nonumber \\
            &  & K_1(\beta(r,m_T)) \cdot  e^{\alpha(r,p_T)  
                       \cdot \cos(\phi_s-\phi_p)} \cdot  \nonumber \\
            &  & e^{-t^2/\tau^2}
\eea
with $\rho(r) = \frac{3}{2}\rho_0\frac{r}{R}$. The system is now confined to a 
cylinder of radius R. There are four parameters: the temperature T, the 
magnitude of the flow transverse boost $\rho_0$, the radius of the cylinder R, 
and the emission duration $\tau$. The parameters that quantify the 
anisotropy of the system ($s_2$ and $\rho_a$) vanish.

  From the phase space density, $f(X,P)$, 
it is straightforward to calculate
transverse mass spectra and source radii. Both the pion source  radii 
and the pion, kaon and proton transverse  mass spectra are well reproduced with 
$\rho _{0} = 0.6$, $T = 110$ MeV, $R = 13$ fm and $\tau =$ 1.5 fm/c. 
It is interesting to notice that
to keep the ratio $R_{out} / R_{side}$ as a function of $p_T$ comparable with 
the values measured by 
STAR\cite{STARHbt} which are close or even below one, the emission duration 
must be on the order of 1 or 2 fm. 
Indeed, since in the usual  Bertsch-Pratt parameterization,
$R_{out}$ is the radius component parallel to the transverse
momentum of the pair ($k_{T}$) while $R_{side}$ is normal to $k_{T}$, the emission
duration increases $R_{out}$ while leaving $R_{side}$ unchanged.
Such a short  emission duration is not currently achieved by 
any realistic microscopic or hydrodynamic model.

The parameters that best fit the identified particle elliptic flow measured by 
STAR~\cite{STARFlow} are $\rho _{0} = 0.61\pm 0.05$, $T = 101\pm 24$
 MeV, $\rho _{a} = 0.04\pm 0.01$ and $s _{2} = 0.04\pm 0.02$. The flow rapidity
and  temperature are consistent with the values obtained from transverse mass 
spectra and pion source size. The system is best described when 
both $\rho_{ a}$ and $s_{2}$  differ from zero, i.e. 
when the system exhibit an asymmetry in momentum and space.  


\section{Pion source geometry with respect to the reaction plane}

  The identified particle elliptic flow favours a system that is asymmetric both
in space and momentum. The pion source should reflect such an asymmetry.
It is thus interesting to study the pion source geometry as a 
function of the angle between the particle momenta and the reaction plane.
 
  Such analysis was performed by the E895 collaboration
~\cite{E895} at the AGS. The important difference between STAR and E895 
is that directed flow ($v_{1}$) cannot yet be measured by STAR, 
which means that only the radii within the transverse plane 
are relevant. The available radii are $R_{out}$, $R_{side}$ and the cross
term $R_{out side}$.  For the theoretical aspect of the analysis see 
~\cite{FlowHbtTh}.

\begin{figure}[ht]
\includegraphics[width=.49\textwidth]{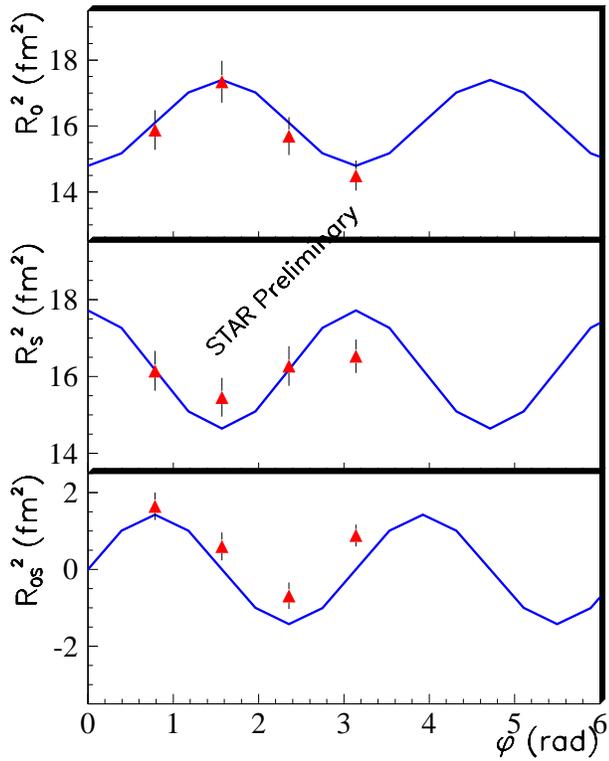}
\caption{\label{HbtFlow} Pion source radii as a function of the 
angle between the pair momenta and the reaction  plane. 
Triangle : data. Solid line : blast wave model calculation.}
\end{figure}

Events were selected using a minimum bias trigger which covers 85\% of the whole
centrality range.  It is mandatory to include peripheral events where the elliptic 
flow is 
maximal. Pion tracks are reconstructed
and identified in the STAR TPC~\cite{STARTpc} in the momentum
range: $0.1<p_{T}<0.6$ GeV/c and $-0.5<Y<0.5$.
The reaction plane is reconstructed as described in ~\cite{STARFlow}.
Pion pairs are sorted into four different bins depending on the angle 
between the pair momentum and the reaction plane.  
 $R_{out}$, $R_{side}$  and $R_{out side}$ are extracted
fitting the four different correlation functions constructed
with each pair sample. Figure ~\ref{HbtFlow}  shows the squared radii $R_{out}^2$, 
$R_{side}^2$ and $R_{out side}^2$ as a function of the mean angle of the 
pairs with respect to the reaction plane.

A clear oscillation of the radii is observed. The line in the figure
is a blast wave model
calculation using the parameters
 $\rho _{0} = 0.6$, $T = 100$ MeV, $\rho _{a} = 0.05$, $s _{2} = 0.05$,  $R=10$ fm,
and $\tau =$ 2 fm/c. These parameters were chosen so that they are consistent 
with the ones extracted 
from identified particle elliptic flow and two-pion interferometry. 
A non zero value of the space asymmetry parameter $s_{2}$ is 
necessary to reproduce the data. When $s_{2}$ is equal to zero the 
momentum asymmetry
expressed by $\rho _{a}$ is not sufficient to reproduce the oscillation of the 
radii. Thus, since  $s _{2}>0$,  this analysis indicates that the pion source 
geometry is an ellipse extended out of the reaction plane. 

\section{Pion-kaon correlation function}

  The correlation of non-identical particles can be used to study the relative 
mean separation between the emission time and/or position of different 
particle species~\cite{NonId, NonId2}. This technique is based on the comparison of two 
different correlation functions.  If one assumes that kaons and pions
are not emitted, on average, at the same time or position, for each pion-kaon
pair there are two configurations: the pion and kaon get closer to each other 
or they move away from one another.  The correlation between the two particles
will be stronger in the first configuration than in the second one. It is
shown in ~\cite{NonId} that these two configurations correspond to 
$ \overrightarrow{v}_{pair} \cdot \overrightarrow{k^{*}}$
either positive or negative ( $\overrightarrow{v}_{pair}$  is the velocity of the 
pair and $\overrightarrow{k^{*}}$ is the
momentum of either the kaon or the pion in the rest frame of the pair).
 Two correlation functions can be constructed :
$C_{+}(\mid \overrightarrow{k^{*}} \mid)$  for pairs with $\overrightarrow{v_{pair}} 
\cdot  \overrightarrow{ k^{*}} >  0$ and $C_{-}(\mid \overrightarrow{k^{*}} 
\mid)$ for $\overrightarrow{v}_{pair} \cdot  \overrightarrow{k^{*}} <  0$.
If both correlation functions are not identical, it implies 
that there is a difference in the average space-time emission
point of pions and kaons, or in other words, there is a space-time
asymmetry in the emission process of pions and kaons. 
Thus, to study whether or not
pions and kaons are emitted at the same time we calculate the ratio 
$C_{-}(k^{*})/C_{+}(k^{*})$ as a function of $k^{*} = \mid \overrightarrow{k^{*}} \mid$.

  In this analysis, only the most central events are 
selected. They represent 12\% of the total hadronic cross section. 
Pions and kaons
are identified by their specific energy loss in the STAR TPC. The 
acceptance of the pions is :
 $0.1<p_{T}<0.6$ GeV/c and $-0.5<Y<0.5$, and $0.3<p_{T}<0.8$ GeV/c
and $-0.5<Y<0.5$ for the kaons. Electrons and positrons are carefully
removed to avoid contamination of the $\pi^{+}-K^{-}$, 
and $\pi^{-}-K^{+}$ pairs by $e^{+}-e^{-}$ pairs which are correlated in a non-trivial
way. Special care is taken to remove two-track merging effect since it 
noticeably influences the $\pi^{+}-K^{+}$ and $\pi^{-}-K^{-}$ correlation functions. 
The purity of the pion and kaon sample needs to be precisely known in order to 
estimate source size parameters from the correlation function. Indeed
misidentified particles, or secondary particles such as pions from $K^{0}_{s}$ 
decay lower the correlation strength. To first order, this effect is equivalent
to increasing the source size which also lowers the average correlation 
strength. The purity analysis is still under way which prevents us 
from making any quantitative statement from this analysis. 

\begin{figure}[ht]
\includegraphics[width=.49\textwidth]{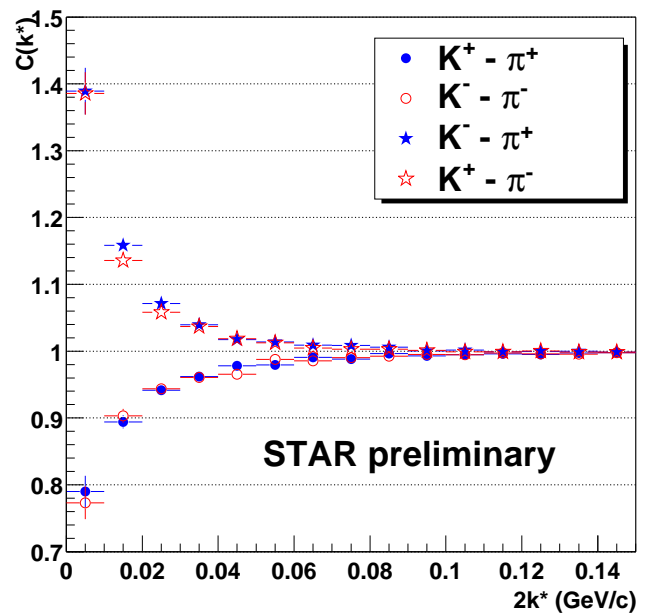}
\caption{\label{PiKCF} $\pi^{+}-K^{+}$, $\pi^{-}-K^{-}$, $\pi^{+}-K^{-}$,
and $\pi^{-}-K^{+}$ correlation functions}
\end{figure}

   The correlation between the pions and kaons arises mainly from their Coulomb
attraction if they have opposite charge or repulsion if they have the same charge.
The $\pi^{+}-K^{+}$, $\pi^{-}-K^{-}$, $\pi^{+}-K^{-}$,
and $\pi^{-}-K^{+}$ correlation functions shown in figure ~\ref{PiKCF} 
exhibit such a behaviour.
A direct comparison to the blast wave model cannot
be performed because these correlation functions depend not only on 
the transverse source size but also on the longitudinal one, which is not
accounted for in the blast wave framework.

\begin{figure}[ht]
\includegraphics[width=.49\textwidth]{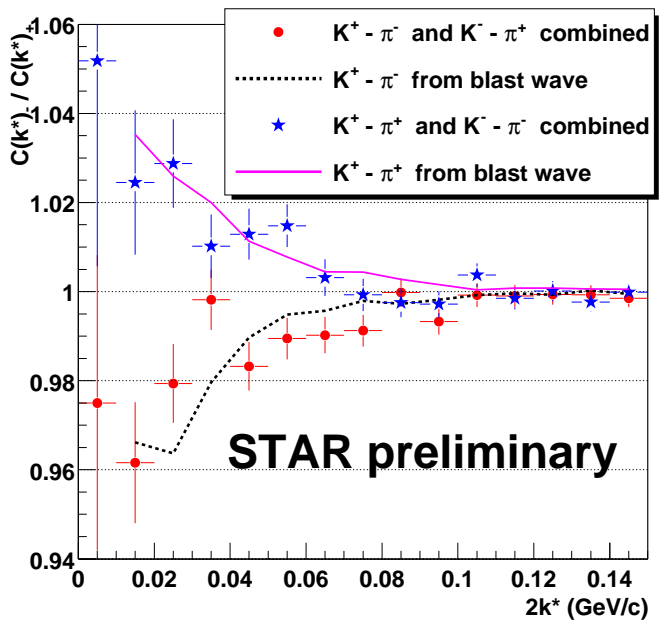}
\caption{\label{PiKRatCF} Ratio of correlation functions $C_{-}(k^{*})/C_{+}(k^{*})$ }
\end{figure}

  The blast wave predictions can however be compared to the ratio of 
correlation functions $C_{-}(k^{*})$ over $C_{+}(k^{*})$ shown in 
figure~\ref{PiKRatCF}. Indeed, the acceptance of each particle
species is symmetric about mid-rapidity, which cancels out
any effect that would be due to a difference in the emission position along the 
longitudinal direction. To improve the statistics, we have combined
the $\pi^{+}-K^{+}$ and $\pi^{-}-K^{-}$ correlation functions and the 
$\pi^{+}-K^{-}$ and $\pi^{-}-K^{+}$  correlation functions. The ratio 
$C_{-}(k^{*})/C_{+}(k^{*})$ is significantly different from unity for both like sign and 
unlike sign pairs which leads us to the conclusion that pions 
and kaons are not emitted at the same space or/and time.

  The extended blast wave model calculations compare well with the data. 
The calculation
was done with $\rho _{0} = 0.6$, $T = 110$ MeV,  $R = 13$ fm, and $\tau=0$ fm/c. 
We showed in section II that, with these 
parameters, the extended blast wave model describes well the transverse mass
spectra and the pion source size. It is striking to notice that the extended blast 
wave model reproduces the data without the need of any extra parameters such 
as one representing the difference between the average emission time of pions 
and kaons.  
Why does the blast wave model predict
an asymmetry between the emission point of the pions and kaons?
To answer this question it is important to notice that at low $k^{*}$, pions
and kaons have the same velocity but not the same momentum. Indeed,
at low $k^{*}$  the pion average $p_{T}$  is equal to 0.1 GeV/c while it is 
equal to 0.4 GeV/c for the kaons. 
In the blast wave model, if the temperature
is equal to zero, particles are emitted at radii that depend only
on their transverse velocity not on their mass. In this case, pions and kaons would
be emitted at the same radii. 
However, switching on the temperature decreases the position-momentum 
correlation, 
which leaves the particles more freedom to fill the cylinder volume.
Thus, the thermal motion pulls the average emission radius away from the
edge of the cylinder, towards the centre of the system. 
Kaons, being less affected by the thermal motion due to their high 
mass compare to the temperature, 
are more likely to be emitted 
close to the edge of the cylinder than pions. In other words, on average, 
kaons are emitted further from the centre of the source than pions which
is why the ratio $C_{-}(k^{*})/C_{+}(k^{*})$ calculated in the extended blast
wave framework, shown on figure ~\ref{PiKRatCF} deviates from unity.

  The interplay between the temperature and transverse flow 
built in the blast wave model introduces
an asymmetry in the emission of the pions and kaons that is in qualitative 
agreement with
the data. 
The non-identical particle correlation 
function can be used to constrain the temperature and flow parameters of the 
extended blast wave model. However, at this stage, given the statistical and 
systematic uncertainties we only conclude
that the blast wave model is in agreement with the data when using the 
same parameters that reproduce the transverse mass spectra and the pion
 source radii.




\section{Conclusions}

   We have shown two new analyses from the STAR experiment at RHIC. 
The analysis
of the pion source geometry with respect to the reaction plane led to the 
conclusion
that the source is an ellipse extended out of plane. The study of the pion-kaon 
correlation functions allows us to state that $<p_{T}> = 0.1$ GeV/c pions and
 $<p_{T}> = 0.4$ GeV/c kaons are not emitted at the same position or/and time.

  We have interpreted our results in the framework of the extended blast wave 
model. We show that the pion source 
geometry, and the pion-kaon space-time emission asymmetry can be described 
by the same set of parameters which are also used to
interpret the transverse mass spectra and the elliptic flow measured by STAR.
The  picture that emerges from this model
is a system with a strong collective expansion that freezes out in a few 
fm/c. Such a scenario is not  currently achieved by any realistic models, 
hydrodynamic or microscopic.



\end{document}